\documentclass[runningheads]{llncs}

\usepackage{epsfig}
\usepackage{multirow}

\begin{document}



\mainmatter

\title{
Comparing and Combining Methods for Automatic Query Expansion
}

\maketitle
\begin{abstract}
Query expansion is a well known method to improve the performance of information 
retrieval systems. In this work we have tested different approaches to extract the candidate query 
terms from the top ranked documents returned by the first-pass retrieval.

One of them is the cooccurrence approach, based on measures of cooccurrence of the candidate and 
the query terms in the retrieved documents. 
The other one, the probabilistic approach, is based on the probability distribution of terms
in the collection and in the top ranked set.

We compare the retrieval improvement achieved by expanding the query with terms obtained with different
methods belonging to both approaches.
Besides, we have developed a na\"ive combination of both kinds of method, with which we have obtained
results that improve those obtained with any of them separately. This result confirms that the information 
provided by each approach is of a different nature and, therefore, can be used in a combined manner.
\end{abstract}




\section{Introduction}

Reformulation of the user queries is a common technique in information retrieval to cover the gap 
between the original user query and his need of information. The most used technique for query 
reformulation is query expansion, where the original user query is expanded with new terms 
extracted from different sources.  Queries submitted by users are usually very short
and query expansion can complete the information need of the users.

A very complete review 
on the classical techniques of query expansion was done by Efthimiadis \cite{Efthi96}. 
The main problem of query expansion is that in some cases the expansion process worsens
the query performance.
Improving the robustness of query expansion has been the goal of many researchers in the last years, 
and most proposed approaches use external collections \cite{Voorhees06,Voorhees05,Voorhees03b}, 
such as the Web documents, to extract candidate terms for the expansion.
There are other methods to extract the candidate terms from the same collection that the search 
is performed on. Some of these methods are 
based on global analysis where the list of candidate terms is generated 
from the whole collection, but they are computationally very expensive and its effectiveness 
is not better than that of methods based on local analysis \cite{QiuF93,qiu95improving,jing94association,Pedersen97}.
We also use the same collection that the search is performed on, but
applying local query expansion, also known as pseudo-feedback or blind feedback, which does not use
the global collection or external sources for the expansion.
This approach was first proposed by Xu and Croft \cite{XuCroft} 
and extracts the expansion terms from the documents retrieved for the original user 
query in a first pass retrieval.

In this work we have tested different approaches to extract the candidate terms from the top ranked 
documents returned by the first-pass retrieval.
There exist two main approaches to rank the terms extracted from the retrieval documents.
One of them is the cooccurrence approach, based on measures of cooccurrence of the candidate and 
the query terms in the retrieved documents. 
The other one is the probabilistic approach, which is based on the differences between the probability 
distribution of terms in the collection and in the top ranked set.
In this paper we are interested in evaluating the different techniques existing to generate the 
candidate term list. Our thesis is that the information obtained with the cooccurrence methods 
is different from the information obtained with probabilistic methods and these two kinds of 
information can be combined to improve the performance of the query expansion process.
Accordingly, our goal has been to compare the performance of the cooccurrence approach and the probabilistic 
techniques and to study the way of combining them so as to improve the query expansion process.

After the term extraction step, the query expansion process requires a further step, namely
to re-compute the weights of the query terms that will be used in
the search process. We present the results of combining different methods for the term extraction and
the reweighting steps.

Two important parameters have to be adjusted for the described process.
One of them is the number of documents retrieved in the first pass to be used for the 
term extraction. The other one is the
number of candidate terms that are finally used to expand the original user query.
We have performed experiments to set both of them to its optimal value in each considered method.


The rest of the paper proceeds as follows:
sections 2 and 3 describe the cooccurrence and probabilistic approaches, respectively;
section 4 presents our proposal to combine both approaches;
section 5 describes the different reweighting methods considered to assign new weights to
the query terms after the expansion process;  
section 6 is devoted to show the experiments performed to evaluate the different expansion 
techniques separately and combined
and section 7 summarizes the main conclusions of this work.

\section{Cooccurrence Methods}

The methods based on term cooccurrence have been used since the 70's to identify some of the 
semantic relationships that exit among terms. In the first works of K. Van Rijsbergen \cite{Rijsbergen} 
we find the idea of using  cooccurrence statistics to detect some kind of semantic 
similarity between terms and exploiting it to expand the user's queries. In fact, this idea is based on the 
Association Hypothesis:

\begin{quotation}
\item
\textit{If an index term is good at discriminating relevant from non-relevant documents then any 
closely associated index term is likely to be good at this.}
\end{quotation}

The main problem with the cooccurrence approach was mentioned by Peat and Willet \cite{PeatW91} 
who claim that  
similar terms identified by cooccurrence tend to occur also very frequently in the collection and 
therefore these terms are not good elements to discriminate between relevant and non-relevant documents. 
This is true when the cooccurrence analysis is done on the whole collection but if we apply it 
only on the top ranked documents discrimination does occur.

For our experiments we have used the well-know Tanimoto, Dice and Cosine coefficients:


\begin{equation}
\rm{Tanimoto}(t_i,t_j) = {c_{ij} \over c_i + c_j -c_{ij}}
\end{equation}

\begin{equation}
\rm{Dice}(t_i,t_j) = {2 c_{ij} \over c_i + c_j}
\end{equation}


\begin{equation}
\rm{Cosine}(t_i,t_j) = {c_{ij} \over \sqrt{c_i c_j}}
\end{equation}
\noindent
where $c_{i}$ and $c_{j}$ are the number of documents in which terms $t_{i}$ and $t_{j}$ occur, 
respectively, and $c_{i,j}$ is the number of documents in which $t_{i}$ and $t_{j}$ cooccur. 

We apply these coefficients to measure the similarity between terms represented by the vectors. 
The result is a ranking of candidate terms where the most useful terms for expansion are at the top.

In the selection method the most likely terms are selected using the equation

\begin{equation}
\rm{rel}(q,t_e) = {\sum_{t_i \in q} q_i {\rm CC}(t_i,t_e)}
\label{ecu_rel}
\end{equation}
\noindent
where {\em CC} is one of the cooccurrence coefficients: Tanimoto, Dice, or Cosine.
Equation \ref{ecu_rel} boosted the terms related with more terms of the original query.

The results obtained with each of these measures, presented in section \ref{sec_experiments},
show that Tanimoto performs better.

\section{Distribution Analysis Approaches}

One of the main approaches to query expansion is based on studying the difference
between the term distribution in the whole collection and in the subsets of documents
that can be relevant for the query. One would expect that terms with little informative content
have a similar distribution in any document of the collection. On the contrary, terms
closely related to those of the original query are expected to be more frequent in the
top ranked set of documents retrieved with the original query than in other
subsets of the collection.  

%
%
%
%
%
%

\subsection{Information-theoretic approach}

One of the most interesting approaches based on term distribution analysis
has been proposed by C. Carpineto et. al. \cite{CarpinetoTOIS01}, and
uses the concept the Kullback-Liebler Divergence \cite{Cover} 
to compute the divergence between the probability distributions of
terms in the whole collection and in the top ranked documents obtained
for a first pass retrieval using the original user query. The most
likely terms to expand the query are those with a high probability in
the top ranked set and low probability in the whole collection.
For the term $t$ this divergence is:

\begin{equation}
KLD_{(PR,PC)}(t) = {P_R(t)  {\rm log} {P_R(t)\over P_C(t)}}
\end{equation}
\noindent
where $P_R(t)$ is the probability of the term $t$ in the top ranked
documents, and $P_C(t)$ is the probability of the term $t$ in the
whole collection.

\subsection{Divergence From Randomness term weighting model}

The Divergence From Randomness (DFR) \cite{amati02} term weighting model
infers the informativeness of a term by the divergence between its distribution in the
top-ranked documents and a random distribution. The most effective DFR term weighting model is
the {\em Bo1 model} that uses the Bose-Einstein statistics \cite{glasgowTrec2004,glasgowTrec2005}:
\begin{equation}
w(t) = {{\rm tf}_x \log_2 ({1+P_n \over P_n}) + \log (1+P_n)}
\end{equation}
where {\em tf}$_x$ is the frequency of the query term in the $x$ top-ranked documents and $P_n$ is given
by $F \over N$, where $F$ is the frequency of the query term in the collection and $N$ is
the number of documents in the collection.

\section{Combined query expansion method}

The two approaches tested in this work can complement each other because they rely on different
information. The performance of the cooccurrence approach  is reduced by 
words which are not stop-words but are very frequent in the collection \cite{PeatW91}. Those words, which
represent a kind of noise, can reach a high position in the term index, 
thus worsening the expansion process. 
However,  precisely because of their high probability in any set of the document collection,
these words tend to have a low score in KLD or Bo1.
Accordingly, combining the cooccurrence measures with others based on the informative content of the terms, such
as KLD or Bo1, helps to eliminate the noisy terms, thus improving the retrieved information 
with the query expansion process.

Our combined model amounts to applying both, a coocurrence method and a 
distributional method and then obtaining the list of candidate  terms
by intersecting the lists provided by each method separately.
Finally, the terms of the resulting list are assigned a new weight by one
of the reweighting method considered.

In the combined approach the number of selected terms depends of the
overlapping between the term sets proposed by both approaches. To
increase the intersection area and obtain enough candidate terms in
the combined list it is neccesary to increase the number of selected
terms for the non-combined approaches. This issue has been studied in
the experiments.


\section{Methods for Reweighting the Expanded Query Terms}

After the list of candidate terms has been generated by one of the methods described above, the selected terms which will 
be added to the query must be re-weighted. Different schemas have been proposed for this task. 
We have compared these schemas and tested which is the most appropriate for each expansion 
method and for our combined query expansion method.

The classical approach to term re-weighting is the Rocchio algorithm \cite{Rocchio}. In this work we have used 
 Rocchio's beta formula, which requires only the $\beta$ parameter, and computes the new weight {\em qtw}
of the term in the query as:
\begin{equation}
\rm{qtw} = {{\rm{qtf} \over \rm{qtf}_{\rm{max}}} + \beta {w(t)\over w_{\rm{max}}(t)}}
\end{equation}
where $w(t)$ is the old weight of term $t$, $w_{\rm{max}}(t)$ is the maximum $w(t)$ of the expanded query terms, 
$\beta$ is a parameter, \rm{qtf} is the frequency of the term $t$ in the query and qtf$_{\rm{max}}$ is the 
maximum term frequency in the query $q$.  In all our experiments, $\beta$ is set to 0.1.

We have also tested other reweighting schemes, each of which directly comes from one of the proposed
methods for the candidate term selection. These schemes use
the ranking values obtained by applying the function defined through each method.
Each of them can only be applied to reweight terms selected with the method it derives from.
This is because these methods require data, collected during the selection process, which are
specific of each of them.

For the case of the reweighting scheme derived from KLD, the new weight is directly obtained applying KLD to
the candidate terms. Terms belonging to the original query maintain their value \cite{CarpinetoTOIS01}.

For the scheme derived from the cooccurrence method, that we called {\em SumCC}, 
the weights of the candidate terms are computed by:
\begin{equation}
\rm{qtw} = {rel(q,t_e) \over \sum_{t_i \in q}q_i}
\end{equation}
where $\sum_{t_i \in q}q_i$ is the sum of the weights of the original terms \cite{zazo2006reina}.

Finally, for the reweighting scheme derived from the Bose-Einstein statistics, a normalization of Bo1 that
we call {\em BoNorm}, we have defined a simple 
function based in the normalization of the values obtained by Bose-Einstein computation:

\begin{equation}
\rm{qtw} = {Bo(t) \over \sum_{t \in cl} Bo(t)}
\end{equation}
where Bo(t) is the Bose-Einstein value for the term $t$, and the sum runs on all terms included in the candidate 
list obtained applying Bose-Einstein statistics.

\section{Experiments}
\label{sec_experiments}

We have used the Vector Space Model implementation provided by Lucene\footnote{http://lucene.apache.org} 
to build our information retrieval system. 
Stemming and stopword removing has been applied in indexing and expansion process.
%
Evaluation is carried out on the Spanish EFE94 corpus, which is part
of the CLEF collection \cite{Peters01} (approximately 215K
documents of 330 average word length and 352K unique index terms)
and on the 2001 Spanish topic set, with 100 topics corresponding to 2001 and 2002 years, of which we only used
the title (of 3.3 average word length).

We have used different measures to evaluate each method. 
Each of them provides a different estimation of the precision of the retrieved documents,
which is the main parameter to optimize when doing query expansion, since recall is always improved
by the query expansion process. The measures considered have been:
\begin{itemize}
\item
MAP {\em (Mean Average Precision)}, which is the average of the precision value (percent of retrieved documents 
that are relevant) obtained for
the top set documents existing after each relevant document is retrieved. 
In this way MAP measures precision at all recall levels and provides a view 
of both aspects.
\item
GMAP, a variant of MAP, that uses a geometric mean rather than an arithmetic mean
to average individual topic results.
\item
Precision@X, which is precision after X
   documents (whether relevant or non-relevant) have been retrieved. If X documents were not retrieved
   for a query, then all missing documents are assumed to be non-relevant.
\item
R-Precision, which measures precision after R documents
   have been retrieved, where R is the total number of relevant documents
   for a query.  
   If R is greater than the number of documents retrieved for a query, then
   the non-retrieved documents are all assumed to be non-relevant.
\end{itemize}

First of all we have tested the different cooccurrence methods described above. 
Table \ref{tco_compara} shows the results obtained for the different measures considered
in this work. We can observe that Tanimoto provides the best results for all the measures, except
for P@10, but in this case the difference with the result of Dice, which is the best,
is very small. According to the results we have selected 
the Tanimoto similarity function as coocurrence method for the rest of the work.
\begin{table}[tbh]
\begin{center}
\begin{tabular}{|l|c|c|c|c|c|}
\hline
               & MAP        &  GMAP       & R-PREC       & P@5       & P@10 \\
\hline
Baseline       &  0.4006    &  0.1941     &    0.4044    &  0.5340   &  0.4670   \\
Cosine         &  0.4698    &  0.2375     &    0.4530    & 0.6020    &  0.5510   \\
Tanimoto       &{\bf 0.4831}&{\bf 0.2464} &{\bf 0.4623}  &{\bf 0.6060}   &  0.5520   \\
Dice           &  0.4772    &  0.2447     &    0.4583    &  0.6020   &{\bf 0.5530}   \\
\hline
\end{tabular}
\end{center}
\caption{Comparing different cooccurrence methods. The Baseline row corresponds to the results
of the query without expansion. P@5 stands for precision after the first five documents retrieved,
P@10 after the first ten, and R-PREC stands for R-precision.
Best results appear in boldface.
}
\label{tco_compara}
\end{table}

\subsection{Selecting the Reweighting Method}

The next set of experiments have had the goal of determining the most appropriate reweighting method
for each candidate term selection method.
Table \ref{rew_meth_coo} shows the results of different reweighting methods (Rocchio and SumCC) applied 
after selecting the candidate terms by the cooccurrence method.
We can observe that the results are quite similar for both reweighting methods,
though Rocchio is slighly better.

\begin{table}[tbh]
\begin{center}
\begin{tabular}{|l|c|c|c|c|c|}
\hline
               & MAP  &  GMAP & R-PREC & P@5 & P@10 \\
\hline
Baseline       &  0.4006    &  0.1941     &    0.4044    &  0.5340       &  0.4670   \\
CooRocchio     &{\bf 0.4831}&{\bf 0.2464} &    0.4623    &  0.6060       &{\bf 0.5520}\\
CooSumCC      &  0.4798    &  0.2386     &{\bf 0.4628}  &{\bf 0.6080}   &  0.5490   \\
\hline
\end{tabular}
\end{center}
\caption{Comparing different reweighting methods for cooccurrence.
{\em CooRocchio} corresponds to using cooccurrence as selection terms method and Rocchio as
reweighting method.
{\em CooSumCC} corresponds to using cooccurrence as selection terms method and 
{\em SumCC} as reweighting method.
Best results appear in boldface.
}
\label{rew_meth_coo}
\end{table}

Table \ref{rew_meth_kld} shows the results of different reweighting methods (Rocchio and kld) applied 
after selecting the candidate terms with KLD.
The best results are obtained using kld as reweighting method.

\begin{table}[tbh]
\begin{center}
\begin{tabular}{|l|c|c|c|c|c|}
\hline
               & MAP  &  GMAP & R-PREC & P@5 & P@10 \\
\hline
Baseline       &  0.4006    &  0.1941     &    0.4044    &  0.5340     &  0.4670   \\
KLDRocchio     &  0.4788    &  0.2370     &    0.4450    &  0.5960     &  0.5480   \\
KLDkld         &{\bf 0.4801}&{\bf 0.2376} &{\bf 0.4526}  &{\bf 0.6080} & {\bf 0.5510}   \\
\hline
\end{tabular}
\end{center}
\caption{Comparing different reweighting methods for KLD.
{\em KLDRocchio} corresponds to using KLD as selection terms method and Rocchio as
reweighting method.
{\em KLDkld} corresponds to using KLD as selection terms method and 
{\em kld} as reweighting method.
Best results appear in boldface.
}
\label{rew_meth_kld}
\end{table}

Table \ref{rew_meth_bo1} shows the results of different reweighting methods (Rocchio and BoNorm) applied 
after selecting the candidate terms with Bo1.
In this case, the best results are obtained using BoNorm as reweighting method.

\begin{table}[tbh]
\begin{center}
\begin{tabular}{|l|c|c|c|c|c|}
\hline
               & MAP  &  GMAP & R-PREC & P@5 & P@10 \\
\hline
Baseline       &  0.4006    &  0.1941     &    0.4044    &  0.5340     &  0.4670     \\
BoRocchio      &  0.4765    &  0.2381     &    0.4450    &  0.5880     &  0.5450     \\
BoBoNorm       &{\bf 0.4778}&{\bf 0.2388} &{\bf 0.4470}  &{\bf 0.5960} &{\bf 0.5470} \\
\hline
\end{tabular}
\end{center}
\caption{Comparing different reweighting methods for Bo1
{\em BoRocchio} corresponds to using Bo1 as selection terms method and Rocchio as
reweighting method.
{\em BoBoNorm} corresponds to using Bo1 as selection terms method and 
{\em BoNorm} as reweighting method.
Best results appear in boldface.
}
\label{rew_meth_bo1}
\end{table}

The results of this section show that the best reweighting method after selecting terms by cooccurrence
is Rocchio, while for the distributional methods in the term selection process, the best
reweighting is obtained with the method derived from themselves,
though Rocchio also provides results very close to the best one.

\begin{figure}[tbp]
\begin{center}
\begin{tabular}{cc}
  \scalebox{0.20}{\epsfig{file=dibujos/coocurrencia-terminos.eps,clip=}} &
  \scalebox{0.20}{\epsfig{file=dibujos/bo-terminos.eps,clip=}}\\
  Cooccurrence & Bose-Einstein statistics\\
  \scalebox{0.20}{\epsfig{file=dibujos/kld-terminos.eps,clip=}} &
  \scalebox{0.20}{\epsfig{file=dibujos/bo-co-terminos.eps,clip=}}\\
  Kullback-Liebler divergence & Combined approach\\
\end{tabular}
\end{center}
\caption[]{Study of the best number of candidate terms to expand the original query with the
different considered methods.
R-PREC stands for R-Precision.
}
\label{fparam_term}
\end{figure}

\subsection{Parameter Study}

We have studied two parameters that are fundamental in query expansion, the number of candidate
terms to expand the query and the number of documents from the top ranked set used to extract
the candidate terms. The optimal value of these parameters can be different for each method, and
thus we have studied them for each case.
The reweighting used for each method has been the one that provides de best results,
and Rocchio for the combined approach.

Figure \ref{fparam_term} shows, for the different expansion methods considered, the MAP and R-PREC 
measures with different numbers of candidate terms to expand the original query.
We can observe that the results of both measures, MAP and R-PREC, indicate similar values,
and that this value is different for each considered method: 
around 25 terms for the cooccurrence method, 40 terms for Bose-Einstein statistics and 
Kullback-Liebler divergence and 75 terms for our combined approach.
The combined approach requires a larger number of selected terms from each basic approach
in order to have enough expansion terms in the intersection list.

Figure \ref{fparam_doc} shows, for the different expansion methods considered, the MAP and R-PREC 
measures with different numbers of documents used to extract the set of candidate query terms.
We can observe that in all cases the best value is around 10 documents.

\begin{figure}[tbp]
\begin{center}
\begin{tabular}{cc}
  \scalebox{0.20}{\epsfig{file=dibujos/co-documents.eps,clip=}} &
  \scalebox{0.20}{\epsfig{file=dibujos/bo1-documents.eps,clip=}}\\
  Cooccurrence & Bose-Einstein statistics\\
  \scalebox{0.20}{\epsfig{file=dibujos/kld-documents.eps,clip=}} &
  \scalebox{0.20}{\epsfig{file=dibujos/bo1-co-documents.eps,clip=}}\\
  Kullback-Liebler divergence & Combined approach\\
\end{tabular}
\end{center}
\caption[]{Study of the best number of documents used to extract the set of candidate query terms.
R-PREC stands for R-Precision.
}
\label{fparam_doc}
\end{figure}

\subsection{Comparing and Combining Both Approaches}

The next step of our experiments has been comparing the overall retrieval performance of
the different expansion method considered, including our combined approach.
The reweighting used for each method has been the one that provides de best results,
and Rocchio for the combined approach.
Table \ref{tmeth_comp_map} shows MAP and GMAP measures, while
table \ref{tmeth_comp_prec} shows R-precision, precision after 5 documents retrieved (P@5) and
after 10 documents (P@10). We can observe that for nearly every measure (except for P@5) 
the best results are obtained by the combination of the Bo1 model with cooccurrence.
The next best result is provided by the other combination considered, KLD with cooccurrence.
These results prove that the information provided by methods belonging to different approaches,
cooccurrence and distributional analysis, is different and thus its combination improves the
results obtained by any of them separately.


\begin{table}[tbh]
\begin{center}
\begin{tabular}{|l|c|c|}
\hline
               & MAP                & GMAP             \\ 
\hline
Baseline       &  0.4006(-)         & 0.1941(-)        \\ 
KLD            &  0.4801(+16.55\%)  & 0.2376(+18.30\%)\\ 
Bo             &  0.4778(+16.15\%)  & 0.2388(+18.71\%)\\ 
Cooccurrence    &  0.4831(+17.07\%)  & 0.2464(+21.22\%)\\ 
BoCo           &  {\bf 0.4964(+19.29\%)}  & {\bf 0.2570(+24.47\%)}\\ 
KLDCo          &  0.4944(+18.97\%)  & 0.2483(+21.82\%)\\ 
\hline
\end{tabular}
\end{center}
\caption{Comparing MAP and GMAP for different methods considered for query expansion.}
\label{tmeth_comp_map}
\end{table}

\begin{table}[tbh]
\begin{center}
\begin{tabular}{|l|c|c|c|}
\hline
               & R-PREC             & P@5               &  P@10\\
\hline
Baseline       & 0.4044(-)          & 0.5340(-)         & 0.4670(-)  \\
KLD            & 0.4526(+10.64\%)     & 0.6080(+12.17\%)  & 0.5510(+15.24\%) \\
Bo             & 0.4470(+9.53\%)     & 0.5960(+10.40\%)    & 0.5470(+14.62\%)  \\
Cooccurrence    & 0.4623(+12.5\%)    & 0.6060(+11.88\%)  & 0.5520(+15.39\%)  \\
BoCo           & {\bf 0.4629(+12.63\%)}   & 0.6220(+14.14\%)  & {\bf 0.5630(+17.05\%)}  \\
KLDCo          & 0.4597(+12.02\%)   & {\bf 0.6240(+14.42\%)}  & 0.5600(+16.60\%)  \\
\hline
\end{tabular}
\end{center}
\caption{Comparing R-Precision (R-PREC), precision after 5 documents retrieved (P@5) and
after 10 documents retrieved (P@10) for different methods considered for query expansion.}
\label{tmeth_comp_prec}
\end{table}

\subsection{Analysis of the results}

We have analyzed the results for some specific queries of our test set.
Table \ref{tmap_queries} compares the MAP measure obtained for cooccurrence, Bo1 and the combination
for the test set queries shown in table \ref{tqueries}.
We can observe that the best result in each case is provided by a different method, 
thus showing that these methods provided different information.
We can also observe that the combined model not always provides the best result. This
suggests to investigate other combination schemes.

\begin{table}
\begin{center}
\begin{tabular}{ll}
\hline
C041 & Pesticidas en alimentos para beb\'es \\

C049 & Ca\'{\i}da de las exportaciones de coches en Jap\'on\\

C053 & Genes y enfermedades \\

C055 & Iniciativa Suiza para los Alpes \\

C058 & Eutanasia \\

C122 & Industria norteamericana del autom\'ovil\\
\hline
\end{tabular}
\caption{Queries used to study the performances of each method in particular cases.}
\label{tqueries}
\end{center}
\end{table}

\begin{table}[tbh]
\begin{center}
\begin{tabular}{|l|c|c|c|c|c|c|}
\hline
Measure         &   41        &   49         &  53      & 55        &  58      &   122\\
\hline
Baseline        &  0.62       & 0.1273       &  0.3098  & 0.2334    &  0.8417  &  0.0760  \\
Cooccurrence    &  0.9428     & 0.2775       &{\bf 0.4901} & 0.6447    &  0.8960  &  0.0588  \\
Bo1             &  0.9270     &{\bf 0.3594}  &  0.4005  & 0.5613    &{\bf 0.9329}  &  0.1130  \\
BoCo            &{\bf 0.9630} & 0.3072       &  0.4724  & {\bf 0.6588} &  0.9223  &{\bf 0.1252}  \\
\hline
\end{tabular}
\end{center}
\caption{Results of the MAP measure for the queries 41, 49, 53, 55, 58, 122.
BoCo stands for the model combining Bo1 and cooccurrence.
The best value appears in boldface.
}
\label{tmap_queries}
\end{table}

\section{Conclusions and Future Works}
We have presented a study of two different approaches, cooccurrence 
and distributional analysis, for query expansion. For each approach we have
considered several models. Results have shown that the query expansion
methodology that we apply is very robust and improves the retrieval results
of the original query with all tested approaches.

The analysis of the results indicates that the statistical information
exploited in each considered approach is different, and this suggests combining
them to improve the results.

We have carried out experiments to measure the improvement of
each method separately, and the combination of them.
Results have shown that a simple combination of the different
query expansion approaches is more efficient than the use of any of
them separately.
This result confirms our thesis that the information exploited by
each approach is different and it is worthwhile to investigate
more sophisticated ways of performing this combination, what we
plan to do in future works.




%
\bibliographystyle{plain}
\bibliography{cicling08}
\nocite{*}
\end{document}